\documentclass{iau}
\usepackage{graphicx}

\title[Structure and kinematics of the Bootes filament] 
{Structure and kinematics\\of the Bootes filament}

\author[O.\,Nasonova, I.\,Karachentsev, \& V.\,Karachentseva]   
{O.\,Nasonova$^1$,\ 
I.\,Karachentsev$^1$,\ 
\and V.\,Karachentseva$^2$}

\affiliation{$^1$Special Astrophysical Observatory of RAS, Nizhnij Arkhyz, Russia\\[\affilskip]
$^2$Main Astronomical Observatory of NASU, Kyiv, Ukraine}

\pubyear{2014}
\volume{308}  
\jname{The Zeldovich Universe: Genesis and Growth of the Cosmic Web}
\editors{A.C. Editor, B.D. Editor \& C.E. Editor, eds.}
\begin{document}

\maketitle

\begin{abstract}
Bootes filament of galaxies is a dispersed chain of groups residing on sky
between the Local Void and the Virgo cluster. We consider a sample of 361
galaxies inside the sky area of $\mathrm{RA} = 13^h\hspace{-0.4em}.\,0 ...
18^h\hspace{-0.4em}.\,5$ and $\mathrm{Dec} = -5^\circ ... +10^\circ$ with radial
velocities $V_{LG} < 2000$~km/s to clarify its structure and kinematics. In this
region, 161 galaxies have individual distance estimates. We use these data to
draw the Hubble relation for galaxy groups, pairs as well as the field galaxies,
and to examine the galaxy distribution on peculiar velocities. Our analysis
exposes the known Virgo-centric infall at $\mathrm{RA} < 14^h$ and some signs of
outflow from the Local Void at $\mathrm{RA} > 17^h$. According to the galaxy
grouping criterion, this complex contains the members of 13 groups, 11 pairs and
140 field galaxies. The most prominent group is dominated by NGC\,5846. The
Bootes filament contains the total stellar mass of $2.7\times10^{12} M_\odot$
and the total virial mass of $9.07\times10^{13} M_\odot$, having the average
density of dark matter to be $\Omega_m = 0.09$, i.e. a factor three lower than
the global cosmic value.

\end{abstract}

\firstsection 
\section{Introduction}

Oservational data on distances and velocities of galaxies in the Local
Supercluster have been enriched significantly by recent optical and HI surveys,
allowing to obtain the nearby field of peculiar motions and hence to study the
local distribution of dark matter. Previously, we considered motions of galaxies
in some sky areas neighbouring the Virgo cluster as the Local Supercluster
centre. In the Virgo Southern Extension filament \cite[(Karachentsev \& Nasonova
2013)]{virsex.13} and the Ursa Majoris cloud \cite[(Karachentsev \etal\
2013)]{ursa.13} we derived a low mean matter density: $\Omega_m=0.11$ and 0.08,
respectively. But in the Coma~I region we suggested the existence of a dark
attractor with the total mass of $\sim2\times 10^{14}M_{\odot}$
\cite[(Karachentsev \etal\ 2011)]{coma.11}.

\section{Observational data}

The initial sample of galaxies was selected from the Lyon Extragalactic Database
= LEDA (http://leda.univ-lyon1.fr) and limited by radial velocities $V_{LG}\leq
2000$~km/s and equatorial coordinates $\mathrm{RA} =
13^h\hspace{-0.4em}.\,0\:...\:18^h\hspace{-0.4em}.\,0$ and $\mathrm{Dec} =
-5^\circ\:... +10^\circ$. The major part of the considered strip is covered by
SDSS survey \cite[(Abazajian \etal\ 2009)]{abaz.06} as well as HIPASS
\cite[(Zwaan \etal 2003)]{zwaan.03} and ALFALFA \cite[(Haynes \etal\
2011)]{hayn.11} HI surveys. It allowed us to perform an independent
morphological classification of galaxies and to determine distances for many
galaxies from the \cite[Tully \& Fisher (1977)]{tf.77} relation between
luminosity of a galaxy and its HI line width $W_{50}$. The resulting list of 361
galaxies includes 161 galaxies with individual distance estimates.

\section{Structure and sub-structures}

The Bootes filament of galaxies is a scattered chain of galaxy groups residing
between the Local Void ($\mathrm{RA}=19^h\hspace{-0.4em}.\,0$,
$\mathrm{Dec}=+3^{\circ}$) and the Virgo cluster
($\mathrm{RA}=12^h\hspace{-0.4em}.\,5$, $\mathrm{Dec}=+12^{\circ}$). The
kinematics of this structure should be influenced both by galaxies infalling
towards the Virgo cluster as well as more eastern galaxies moving away from the
expanding Local Void \cite[(Nasonova \& Karachentsev 2011)]{void.11}.

\begin{figure}[h]
\begin{center}
\includegraphics[height=\textwidth,keepaspectratio,angle=270]{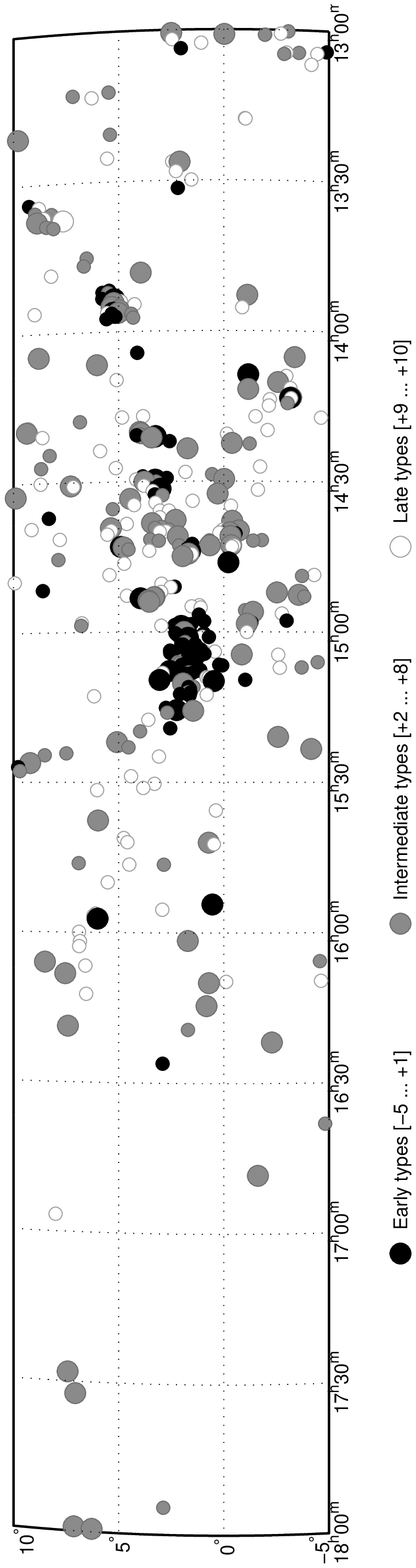}\par
\vspace{0.3cm}
\includegraphics[width=\textwidth,keepaspectratio]{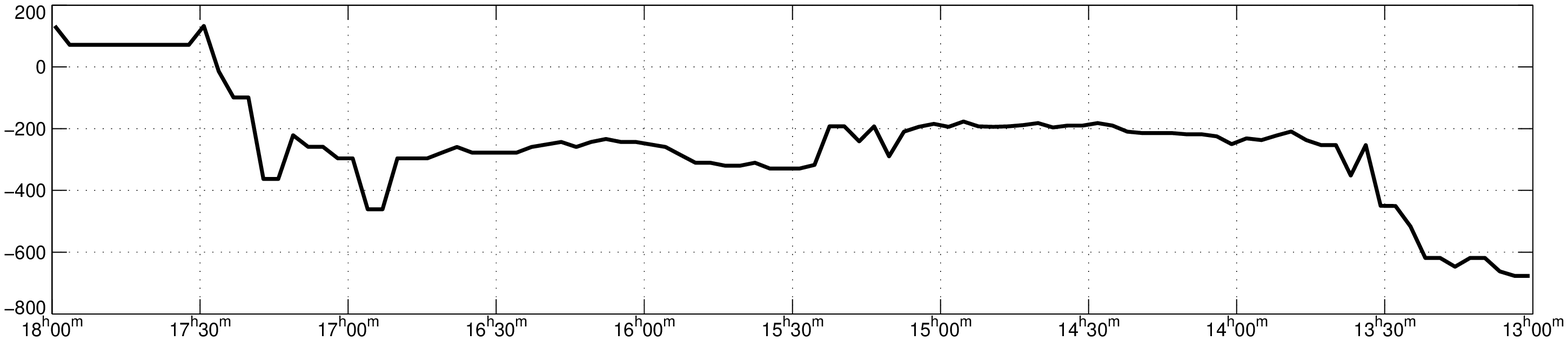}\par
\caption{Upper panel: distribution of galaxies in the Bootes strip. Bottom
panel: a poliline represents the running median of peculiar velocity with a
window of $0^h\hspace{-0.4em}.\,5$.}
\label{fig2}
\end{center}
\end{figure}

The upper panel of Fig.~1 presents the distribution of 361 galaxies in the
Bootes strip. The galaxies of different morphological types are shown by circles
of different density. According to the galaxy grouping criterion \cite[(Makarov
\& Karachentsev 2011)] {mk.11}, the Bootes strip contains 13 groups, 11 pairs
and 140 field galaxies. The most notable feature in the strip is the compact
group around NGC\,5846 (just right from the strip center), which numbers 74
members with measured radial velocities. The bottom panel of Fig.~1 shows the
running median of the peculiar velocity, $V_{pec} = V_{LG} - 72 D_{Mpc}$, with a
window of $0^h\hspace{-0.4em}.\,5$ along RA. The most common value is $V_{pec}=
-250$~km/s, which remains almost flat from $14^h$ to $17^h$. Galaxies in the
Virgo infall zone ( $\mathrm{RA} < 14^h$) demonstrate clearly a droop of the
$V_{pec}$ median. To the contrary, in the vicinity of the Local Void the median
rises which is quite expectable since galaxies move away from the void centre.

\section{Local density of matter}

The virial (projected) mass distribution of galaxy groups (squares) and pairs 
(triangles) in the considered strip versus their total stellar mass is depicted
in Fig.~2. There is a positive correlation between virial and stellar masses,
well-known from other data. While masses are small, the significant vertical
scatter is caused mainly by projection factors.

\begin{figure}[h]
\begin{center}
\includegraphics[height=0.60\textwidth,keepaspectratio,angle=270]{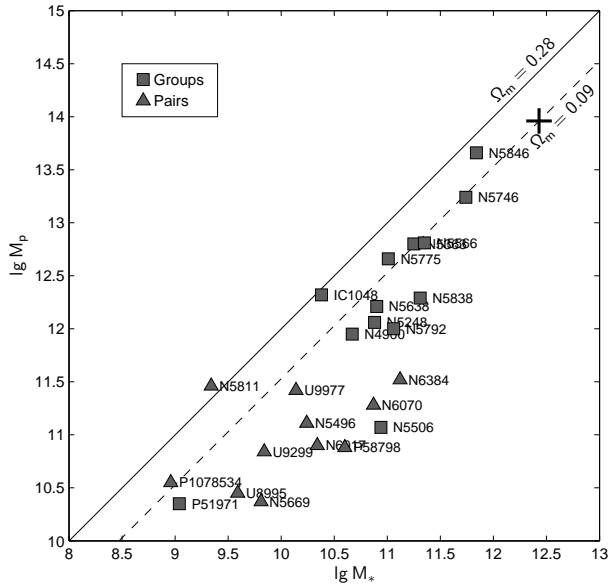}\par
\caption{The relation between projected (virial) mass and total stellar mass.}
\label{fig3}
\end{center}
\end{figure}

According to Jones et al (2006), the mean density of stellar matter in the
universe is $4.6\times 10^8 M_\odot/\mathrm{Mpc}^3$. The global matter density
$\Omega_m = 0.28$ is equivalent to dark-to-stellar matter ratio of
$M_{DM}/M_*=97$. This value is shown as a diagonal in Fig.~2. All the groups and
all the pairs except one in the Bootes strip are situated under this line. The
Bootes filament contains the total stellar mass of $2.7\times10^{12}M_\odot$ and
the total virial mass of $9.07\times10^{13} M_\odot$, having the average ratio
$\sum M_p/\sum M_*=33$ or the local density of dark matter $\Omega_m = 0.09$.
This is a factor three lower than the global cosmic value. The sum of virial
masses-to-sum of stellar masses ratio for all the groups and pairs is plotted in
Fig.~2 as a cross. The dashed line drawn through the cross indicates the mean
mass density $\Omega_m$ (Bootes) $\simeq 0.09$. Considering field galaxies
should only reduce slightly this proportion, as they contribute evidently both
to numerator and denominator of the ratio $\sum M_p/\sum M_*$. Thus, the
observational data on galaxy motions in the Bootes strip give us an argument
that this filamentary structure does not contain a large amount of dark matter.
This statement is based on internal (virial) motions of galaxies, but supposing
that 3--5 times larger mass is hidden in the Bootes filament between the groups,
then the velocity dispersion for centres of groups and pairs should be
considerably larger than what is observed.

\vspace{0.18cm}

{\scriptsize{}%
This work is supported by the Russian Science Foundation (project No.
14-12-00965), Russian Foundation for Basic Research (grant No. 13-02-90407) and
the State Fund for Fundamental Researches of Ukraine (grant No. F53.2/15). Olga
Nasonova thanks the non-profit Dmitry Zimin’s Dynasty Foundation for the
financial support. This research has made use of NASA/IPAC Extragalactic
Database (http://ned.ipac.caltech.edu), HyperLeda database
(http://leda.univ-lyon1.fr) and SDSS archive (http://www.sdss.org).





\end{document}